\documentstyle[12pt,aaspp4]{article}
\tighten
\eqsecnum
\received{}
\accepted{}
\journalid{}{}
\articleid{}{}
\slugcomment{To appear in {\it The Astronomical Journal} (Sept. 1999)}
\begin{document}
 
\title{A Search for Radio Emission at the Bottom of the Main Sequence and
Beyond}

\author{Anita Krishnamurthi, Giuseppe Leto\altaffilmark{1}, and Jeffrey
L. Linsky} \affil{JILA, University of Colorado and National Institute of
Standards and Technology, Boulder, CO 80309-0440\\ email:
anitak@casa.colorado.edu; gleto@casa.colorado.edu;
jlinsky@jilau1.colorado.edu}

\altaffiltext{1}{Visiting from: OACt, Catania Astrophysical Observatory,
Viale A. Doria 5, 95125 Catania, Italy; email: gleto@ct.astro.it}

\begin{abstract}

We have used the VLA to conduct a deep search for 3.6 cm radio emission
from nearby very low mass stars and brown dwarfs.  The G\"udel-Benz
relation is used to predict radio luminosities for some very low mass
stars and candidate brown dwarfs with measured X-ray fluxes.  The
predicted radio fluxes are quite small, whereas the measured radio flux
from the brown dwarf candidate $\rho$ Oph GY 31 is relatively strong.
In light of our new observations, this object remains an anomaly.  We
present upper limits for our measured radio fluxes at 3.6 cm for our
targets.

{\it Keywords :} stars: activity -- stars: low--mass, brown dwarfs --
stars: magnetic fields
 
\end{abstract}
 
\section{INTRODUCTION}

Radio emission is a common observational indicator of stellar magnetic
activity.  Late--M dwarfs like UV Cet and Proxima Cen emit quasi-steady
continuum radio emission at centimeter wavelengths, which has been
interpreted as gyro-synchrotron emission or gyro-resonance emission
(e.g., \cite{linsky83}; \cite{gudel96}).  This emission could be the
superposition of many small flares.  These stars occasionally show large
flares at radio wavelengths that may be coherent emission or
gyro-synchrotron emission from highly energetic electrons.  Flares have
been detected on some very late type stars such as VB 10 (M8e) and 2MASS
J0149090+295613 (M9.5V) in the optical, X-ray and ultraviolet (e.g.,
\cite{linsky95}; \cite{liebert99}).

While the nature of the dynamo mechanism for amplifying magnetic fields
in fully convective stars (those later than about spectral type M5
corresponding to $M < 0.3M_{\odot}$) is probably different from the more
massive stars with radiative cores, Fleming et al. (1993) and others
find no observed change in activity indicators such as X-ray surface
fluxes near spectral type M5.  No qualitative change in the dynamo is
expected at the boundary dividing the lowest mass stars from the
substellar objects since both types of objects are fully convective, but
this assumption has not been tested.  Studies of stellar activity in the
lowest mass M dwarfs have shown that these stars are very active X-ray
emitters (e.g., \cite{fleming93}, \cite{basri95}, \cite{drake96},
\cite{giampapa96}).

Brown dwarfs are very low mass objects (M $<$ 0.08 M$_{\odot}$) with
cores that are not hot enough to support hydrogen burning.  They can
only burn deuterium very early in their evolution.  The long search for
these substellar objects has recently uncovered brown dwarfs in young
open clusters, as free-floating objects in the field, and around nearby
M dwarfs (c.f. \cite{nakajima95}; \cite{zap97}; \cite{tinney97};
\cite{ruiz97}).  Several ongoing surveys, including the DENIS and 2MASS
surveys, are currently discovering these objects (\cite{delfosse99};
\cite{kirk99}).

It is commonly believed that brown dwarfs should have magnetic fields,
flares, nonthermal particles, and nonthermal radio emission like the
very late-M dwarfs because (i) both late-M dwarfs and brown dwarfs are
fully convective with similar interior structures and (ii) brown dwarfs
appear to be rapid rotators (e.g., \cite{basri95}) that should
strengthen the dynamo mechanism.  Therefore, brown dwarfs should have
magnetic dynamos.  The detection of either steady or flare nonthermal
radio emission from brown dwarfs would provide unambiguous evidence that
brown dwarfs do indeed have magnetic fields and the related phenomena
that magnetic fields produce in stars.  The detection of magnetic fields
in brown dwarfs would confirm the prediction that the magnetic behavior
of normal stars continues down to substellar masses, and thus be of
great significance in our understanding of brown dwarfs.

The VLA has been used previously to observe extremely low mass stars,
including the confirmed brown dwarf, Gl~229B, but no detections have
been reported.  There is one detection of relatively strong (but highly
variable) radio emission from a brown dwarf candidate: $\rho$ Oph GY 31
shows 3.6 cm radio emission varying from 0.5 mJy to $\le$0.1 mJy
(\cite{neuhauser99}).

In this paper, we report the results of an observational program to look
for radio emission from some very low mass stars and brown dwarfs using
the VLA.$^{1}$\footnotetext[1]{The National Radio Astronomy Observatory
is a facility of the National Science Foundation operated under
cooperative agreement by Associated Universities, Inc.}  We conducted a
deep search at X-band (corresponding to a wavelength of 3.6 cm or a
frequency of 8.46 GHz) where the VLA is most sensitive.  The previous
observations were mostly at longer wavelengths and for short integration
times.  Sec. 2 describes our observations and data reduction methods,
Sec. 3 presents our results, while Sec. 4 summarizes our conclusions.

\section{OBSERVATIONS AND DATA REDUCTION}

We observed four very low-mass stars (spectral types M7--M9) and two
brown dwarfs to study radio emission as we proceed from the lowest mass
stars to substellar objects.  We selected targets that lie within 12 pc
of the Sun to maximize chances of detection.  Some characteristics of
our targets are presented in Table 1.  As the ages of these objects are
not known accurately, we list in column 6 ages estimates quoted in the
literature.

The brown dwarf Kelu--1 has not previously been observed at radio
frequencies.  The other targets in our sample were observed by the VLA,
mostly at wavelengths where the VLA is less sensitive, or for shorter
integration times.  None of these previous observations have yielded any
detections.

In order to increase the chances of detection, we used the maximum
allowed bandwidth (50 MHz) for each of the two observing frequencies
inside the X band, for a total of 100 MHz bandwidth.  The observations
were made from UT 14:00 to UT 24:00 on JD 2451109.5 (1998 Oct. 23) and
from UT 11:30 to UT 21:30 on JD 2451110.5 (1998 Oct. 24).  The exact
on-source integration times are presented in Table 2.

The VLA was in MOVE configuration (B$\rightarrow$C) during our
observations.  We determined that the northern arm of the array was in B
configuration, while the other 2 arms had already been moved to the C
configuration.  The resulting measured beam width was 2.4$\arcsec$,
which is very close to the beam width of 2.3$\arcsec$ for the C
configuration of the array at X band.

For the flux density calibration, we chose the compact radio source 3C286
(1331+305). The phase calibrators for each of the sources were selected
from the VLA calibrator manual to lie within 5$^{\circ}$ of the source.
The measured flux densities are close to the values reported in the VLA
calibrator manual (\cite{vlacal}).  Our observing sequence included the
primary flux density calibrator observation for about 10 minutes, and an
observation of the phase calibrators at least every 30 minutes.

For the data reduction, we used the Astronomical Image Processing System
(AIPS, release 15Apr1998) developed at National Radio Astronomy
Observatory (NRAO).  All necessary steps of the standard reduction
pipeline to get the total intensity map were done using AIPS tasks.  The
final cleaning was performed using the Cotton-Schwab algorithm.  Little
data editing was necessary due to an overall good data quality despite
the uncertainties in the antenna positioning that could be introduced by
the ``moving'' configuration.

We evaluated the r.m.s. noise levels (listed in Table 2) by selecting
large areas on the maps where no sources appear.  Assuming that the
system temperature of the antennas and the efficiencies are equal to the
tabulated values, we calculated the expected r.m.s. noise for an
integration time, t$_{int}$ = 2.5 h (150 minutes).  Using the parameters
for our observational setup, the calculated r.m.s. noise is
$\sim$10$\mu$Jy, which is close to the actual values we have measured
from our constructed maps.

\section{RESULTS AND DISCUSSION}

We did not detect our targets at 3.6 cm or see any radio flares in our
radio maps, despite the long integration times.  We present our measured
flux upper limits in Table 2.  The integration times for our targets
were selected to provide a detection of the target if it were as radio
luminous at 3.6 cm as the relatively inactive M5.5 V star, UV Cet (flux
of 1 mJy at a distance of 2.7 pc, \cite{gudel96}).  Our observations
allowed us to reach the sensitivity level needed to detect our stars if
the scaling from UV Cet is realistic.

The computed L$_{3.6 cm}$/L$_{bol}$ upper limit ratios for our observed
targets are presented in column 5 of Table 2.  This ratio is calculated
to be 1.7 $\times10^{-18}$ for UV Ceti.  We see that the upper limit for
the ratio of radio luminosity at 3.6 cm to the bolometric luminosity for
our targets lies between 60 times larger and three times smaller than
the ratio for UV Ceti.

A source was detected in the map made for the VB~10 (Gl~752B) field
(Table 3).  The source is a 20$\sigma$ detection, and thus is real.
Gl~752 is a binary system with the primary being a high proper motion
M2.5 star.  We checked to see if our detected source corresponds to the
position of the primary, Gl~752A, and determined that it did not.  The
source was previously detected by Leto et al. (in preparation).  It is
interesting to note that Leto et al. had radio images of this field at
Q-band (7mm), K-band (1.3cm), U-band (2cm) and X-band (3.6cm).  However,
they detected a source at the same position as reported in this paper
only in the X-band.  This source has also been cataloged in the NRAO VLA
sky survey at 1.4 GHz with a measured flux of 11.4 mJy (\cite{condon98}
\footnote{A brief description of the catalog is available at
``http://www.cv.nrao.edu/$\sim$jcondon/nvss.html''.  A search of this
catalog may be conducted through the web interface at
``http://www.cv.nrao.edu/NVSS/NVSS.html''}).  We would like to point out
that the flux levels are different in the two observations of this
object at 3.6 cm (Leto et al., in preparation; present paper).  This
might imply that the source is variable.

The detection of activity indicators such as radio emission, X-ray
emission, and starspots on these very low mass stars and brown dwarfs
would indicate the presence of magnetic fields on these objects.  There
have been some optical monitoring campaigns to determine rotational
periods for brown dwarfs by measuring spot modulation (e.g.,
\cite{dt99}).  The optical campaigns have yielded periods for very low
mass stars close to the substellar boundary, although there are still no
measurements of photometric rotation periods for bona-fide brown dwarfs.
Highly variable radio emission at 3.6 cm has been detected from the
young brown dwarf candidate $\rho$ Oph GY 31, which indicates the
presence of a magnetic field (\cite{neuhauser99}).  X-rays have recently
been detected from some other very young candidate and bona-fide brown
dwarfs (\cite{briceno98}; \cite{neuhauser98}; \cite{neuhauser99}), but
no signs of magnetic activity have been detected from bona-fide brown
dwarfs older than a few million years.  A recent paper by Neuh\"{a}user
et al. (1999) discusses a possible explanation: as brown dwarfs age,
they cease to burn deuterium in their cores, their cores cool off, and
the decreased temperature difference between the core and the surface
leads to less vigorous convection.  As strong convection is likely
needed to make the dynamo work, rapidly rotating older brown dwarfs with
weak convection may not have strong enough magnetic fields to produce
strong X-ray emission or nonthermal radio emission or show significant
spot coverage.

We have predicted the radio flux for the bona-fide brown dwarfs and
brown dwarf candidates that have been detected in X-rays, using the
G\"{u}del and Benz (1993) relation $L_{X}=10^{15.5} \times L_{R}$, where
$L_{R}$ is the radio flux at 3.6 or 6 cm.  This relation is based on
X-ray and radio observations of many active stars covering a broad range
of spectral types, but not brown dwarfs.  Table 4 presents the predicted
radio fluxes for these objects as well as for the very low mass M dwarfs
VB 8 and VB 10.  We see that the predicted fluxes from the brown dwarfs
are very low, less than 1$\mu$Jy.  In particular, the predicted flux for
VB 8 is close to our measured upper limit, while the predicted flux for
VB 10 is much lower than our measured upper limit.

We also predicted the radio flux for GY 31, the candidate brown dwarf in
Ophiucus.  This object has very low level X-ray emission
($log(L_{X}/L_{bol}$) = -5.81) and 3.6 cm radio emission varying from
0.5 mJy in its high state to $\le$0.1 mJy in its low state
(\cite{neuhauser99}).  Applying the G\"{u}del-Benz relation to this
star, we find that its predicted radio emission, based on its X-ray
emission would be $\sim$0.05$\mu$Jy.  This is four orders of magnitude
below the observed value.  Hence, we speculate that this relation may
not be valid at the very bottom of the main-sequence and in the
substellar regime due to differences in convection velocities and
possibly weaker dynamos (and thus magnetic fields). It is also possible
that the radio flux for GY 31 was measured during an active period (such
as a flare).

Thus, with exceptions such as GY 31, the radio fluxes for very low mass
objects older than a couple of million years maybe too low to be
detectable with current technology and sensitivity limits.

\section{CONCLUSIONS}

We observed four very low-mass M dwarfs (M7--M9) and two brown dwarfs
within 12 pc of the Sun with the VLA at 3.6 cm.  Our long integrations
were designed to detect radio emission from brown dwarfs and to
establish a continuity of this emission from the bottom of the main
sequence to the substellar objects.  However, our observations did not
yield a detection of any of the sources.  We present some upper limits
of detection for these objects.

The detection of either steady or flare nonthermal radio emission from
brown dwarfs would provide unambiguous evidence that brown dwarfs indeed
do have magnetic fields and the phenomena that magnetic fields produce
in stars.  Brown dwarfs are probably most magnetically active when they
are young, (of the order of a few million years).  Both Kelu--1 and
Gl229B are relatively old sources ($\sim$1 Gyr and 10 Gyr respectively)
as are the dMe stars we studied (several times 10$^{8}$ yrs).  Many of
the young brown dwarfs that have been discovered recently are in distant
clusters.  Our calculations show that their radio fluxes may be too low
to be detected with current sensitivity limits.  A deep search for radio
emission from a nearby young brown dwarf, if one is discovered, should
prove more successful.  The search for brown dwarfs is currently very
vigorous and we will soon have such candidates to study.  A detection of
radio emission from these enigmatic objects would allow placement of
better constraints on the hot plasma in the coronae of these objects and
the mechanisms by which it may be heated.

\acknowledgements
This work was supported by NASA grant H-04630D to the University of
Colorado.

\begin{deluxetable}{lllcclc}
\tablenum{1}
\tablewidth{0pt}
\tablecaption{Targets}
\tablehead{
\colhead{Name} & 
\colhead{RA (2000)}& 
\colhead{DEC (2000)} &
\colhead{Spectral Type} & 
\colhead{Dist. (pc)} &
\colhead{Age (Gyr)} &
\colhead{Ref.}
}
\startdata
Gl 229B	& 06 10 35.07 & -21 51 17.6 & BD~~ & ~~5.7	& 10 Gyr (?) & a\\
LHS 2065& 08 53 38.0  & -03 29.3    & M9e  & ~~8.5 	&several$\times$10$^{8}$yr & b\\
Kelu 1	& 13 05 40.2  & -25 41 06   & BD~~ & 10.0 	& $\sim$1 Gyr  & c\\
Gl 569B	& 14 54 28.0  & +16 06.2    & M8.5 & 10.5	&several$\times$10$^{8}$yr & d\\
VB 8	& 16 55 38.33 & -08 22 51.9 & M7e  & ~~6.5	&several$\times$10$^{8}$yr & b\\
VB 10	& 19 16 59.90 & +05 10 18.4 & M8e  & ~~5.9	&several$\times$10$^{8}$yr & b\\
\enddata
\tablenotetext{a}{Allard et al. 1996}
\tablenotetext{b}{Reid, Tinney, \& Mould 1995}
\tablenotetext{c}{Ruiz, Leggett, \& Allard 1997}
\tablenotetext{d}{Forrest, Skrutskie, \& Shure 1988}
\end{deluxetable}

\small
\begin{deluxetable}{lccccc}
\tablenum{2}
\tablewidth{0pt}
\tablecaption{Measurements made}
\tablehead{
\colhead{Name} & 
\colhead{Integration time} &
\colhead{RMS}& 
\colhead{3$\sigma$ flux upper limit} &
\colhead{$L_{3.6}/L_{bol}$} &
\colhead{Flux ratio to UV Ceti} \\
\colhead{ } & \colhead{(minutes)} & \colhead{($\mu$Jy/beam)} &
\colhead{S($\mu$Jy)} &  &\colhead{at same distance} }
\startdata
Gl 229B	 & 195.45 & 23.4 & 70.2	& $<$ 1.1$\times10^{-16}$    & $<$ 0.36 \\
LHS 2065 & 146.02 & 16.3 & 48.9	& $<$ 4.16$\times10^{-18}$   & $<$ 0.80 \\
Kelu--1	 & 209.20 & ~~9.2& 27.6	& $<$ 2.8$\times10^{-17}$    & $<$ 0.77 \\
Gl 569B	 & 146.77 & 10.3 & 30.9	& $<$ 3.1$\times10^{-18}$    & $<$ 0.79 \\
VB 8	 & 175.05 & ~~7.5& 22.5	& $<$ 5.73$\times10^{-19}$   & $<$ 0.22 \\
VB 10	 & 146.58 & 26.6 & 79.8	& $<$ 2.26$\times10^{-18}$   & $<$ 0.53 \\
\enddata
\end{deluxetable}
\normalsize

\begin{deluxetable}{lcccc}
\tablenum{3}
\tablewidth{0pt}
\tablecaption{Source in VB10 field}
\tablehead{
\colhead{RA (2000)} & 
\colhead{Dec. (2000)}& 
\colhead{RMS}&
\colhead{Source Flux}&
\colhead{Previously measured flux} \\
\colhead{ } & \colhead{ } & \colhead{($\mu$Jy/beam)} &\colhead{(mJy)} &
\colhead{(Leto et al. in prep.) (mJy)}
}
\startdata
19 16 56.66 & +05 11 27.20 & 26.6 & 0.516 & 1.8 \\
\enddata
\end{deluxetable}

\begin{deluxetable}{lcccc}
\tablenum{4}
\tablewidth{0pt}
\tablecaption{Predicted radio luminosities}
\tablehead{
\colhead{Name} & 
\colhead{log L$_{X}$} &
\colhead{L$_{R}$ (ergs/s/Hz)} &
\colhead{d (pc)} &
\colhead{S$_{3.6}(\mu$Jy)} \\
\colhead{ } & \colhead{ } & \colhead{[predicted as
$\frac{L_{X}}{10^{15.5}}$]} &\colhead{ }  & \colhead{[from predicted
L$_{R}$]}
}
\startdata

Cha H$\alpha$ I & 28.41 & 8.13$\times10^{12}$ & 160 & 0.26 \\
GY 202 		& 28.19 & 4.9$\times10^{12}$  & 160 & 0.16 \\
V410 X-ray 3    & 28.68 & 1.51$\times10^{13}$ & 140 & 0.64 \\
V410 Anon 13    & 28.26 & 5.76$\times10^{12}$ & 140 & 0.24 \\
Tau MHO-4       & 28.87 & 2.34$\times10^{13}$ & 140 & 0.99 \\
Tau MHO-5       & 28.17 & 4.68$\times10^{12}$ & 140 & 0.2~~  \\
VB 10 		& 26.34	& 6.92$\times10^{10}$ & 5.74& 1.74 \\
VB 8 		& 27.54 & 1.1$\times10^{12}$  & 6.4 & 22 \\

\enddata
\end{deluxetable}


\begin{thebibliography}{}

\bibitem[Allard et al. 1996]{all96} Allard, A., Hauschildt, P.H.,
Baraffe, I., \& Chabrier, G. 1996, ApJ, 465, L123

\bibitem[Basri \& Marcy 1995]{basri95} Basri, G., \& Marcy, G. 1995, AJ,
109, 762 

\bibitem[Brice\~{n}o et al. 1998]{briceno98} Brice\~{n}o, C., Hartmann,
L.W., Stauffer, J.R., Mart\'{i}n, E.L. 1998, AJ 115,2074

\bibitem[Condon et al. 1998]{condon98} Condon et al, 1998, AJ, 115, 1693

\bibitem[Delfosse et al. 1999]{delfosse99} Delfosse, X., Tinney, C.G.,
Forveille, T., Epchtein, N., Borsenberger, J., Fouque, P., Kimeswenger,
S., \& Tiphene, D. 1999, A\&AS, 135, 41

\bibitem[Drake et al. 1996]{drake96} Drake, J.J., Stern, R.A.,
Stringfellow, G.S., Mathioudakis, M., Laming, J.M., \& Lambert,
D.L. 1996, ApJ, 469, 828

\bibitem[Fleming et al. 1993]{fleming93} Fleming, T.A., Giampapa,
M.S., Schmitt, J.H.M.M., \& Bookbinder, J.A. 1993, ApJ, 410, 387 

\bibitem[Forrest, Skrutskie, \& Shure 1988]{forrest88} Forrest, W.J.,
Skrutskie, M.F., \& Shure, M. 1988, ApJ, 330, L119

\bibitem[Giampapa et al. 1996]{giampapa96} Giampapa, M.S., Rosner, R.,
Kashyap, V., Fleming, T.A., Schmitt, J.H.M.M., \& Bookbinder,
J.A. 1996, ApJ, 463, 707

\bibitem[G\"{u}del \& Benz 1993]{gudel93} G\"{u}del, M. \&
Benz, A.O. 1993, ApJ, 405, L63

\bibitem[G\"{u}del \& Benz 1996]{gudel96} G\"{u}del, M. \& Benz,
A.O. 1996, {\it "Radio Emission from the Stars and the Sun"},
Eds. A.R. Taylor and J.M. Paredes, APS Conf. Ser. 93, 303

\bibitem[Kirkpatrick et al. 1999]{kirk99} Kirkpatrick, J.D., Reid,
I.N., Liebert, J., Cutri, R.M., Nelson, B., Beichman, C.A., Dahn,
C.C., Monet, D.G., Gizis, J.E., \& Skrutskie, M.F. 1999, submitted

\bibitem[Leto et al., in preparation]{leto99} Leto, G. et al. 1999, in
preparation 

\bibitem[Liebert et al. 1999]{liebert99} Liebert, J., Kirkpatrick, J.D.,
Reid, I.N., \& Fisher, M.D. 1999, ApJ, in press

\bibitem[Linsky \& Gary, 1983]{linsky83} Linsky, J.L., \& Gary,
D.E. 1983, \apj, 274, 776

\bibitem[Linsky et al. 1995]{linsky95} Linsky, J.L., Wood, B.E.,
Brown, A., Giampapa, M.S., \& Ambruster, C. 1995, \apj, 455, 670

\bibitem[Nakajima et al. 1995]{nakajima95} Nakajima, T., Oppenheimer,
B.R., Kulkarni, S.R., Golimowski, D.A., Matthews, K., \& Durrance,
S.T. 1995, Nature, 378, 463 

\bibitem[Neuh\"{a}user \& Comer\'{o}n 1998]{neuhauser98} Neuh\"{a}user
R. \& Comer\'{o}n F. 1998, Science, 282, 83

\bibitem[Neuh\"{a}user et al. 1999]{neuhauser99}  Neuh\"{a}user et al.
1999, A\&A, 343, 883

\bibitem[Perley \& Taylor 1999]{vlacal} Perley, R.A. \& Taylor, C.B.
1999, VLA Calibrator manual

\bibitem[Reid, Tinney, \& Mould 1994]{reid94} Reid, N., Tinney, C.G., \&
Mould, J. 1994, AJ, 108, 1456

\bibitem[Ruiz, Leggett, \& Allard 1997]{ruiz97} Ruiz, M.T., Leggett,
S.K., \& Allard, F. 1997, ApJ, 491, L107 

\bibitem[Terndrup et al. 1999]{dt99} Terndrup, D.M., Krishnamurthi, A.,
Pinsonneault, M.H., \& Stauffer, J. 1999, in preparation

\bibitem[Tinney, Delfosse, \& Forveille 1997]{tinney97} Tinney, C.G.,
Delfosse, X., \& Forveille, T. 1997, ApJ, 490, L95

\bibitem[Zapatero-Osorio et al. 1997]{zap97} Zapatero-Osorio, M.R.,
Rebolo, R., Martin, E.L., Basri, G., Magazzu, A., Hodgkin, S.T.,
Jameson, R.F., \& Cossburn, M.R. 1997, ApJ, 491, L81

\end{thebibliography}
\end{document}